\documentclass[prl,twocolumn,superscriptaddress,floatfix]{revtex4-2}
\usepackage{graphicx}
\usepackage{latexsym}
\usepackage{amsmath}
\usepackage{amsfonts}
\usepackage{amssymb}
\usepackage{bm}
\usepackage{txfonts}
\newcommand{\fig}[2]{\includegraphics[width=#1]{#2}}
\def\bk{{\textbf{k}}}
\def\bQ{{\textbf{Q}}}

\def\mT{{{\mathcal{T}}}}

\begin{document}
\title{Engineering tunable $\bm p$-wave magnetism in antiferromagnetic bilayers}

\author{Yu-Han Lin}
\affiliation{College of Mathematics and Physics, Ningde Normal University, Ningde 352100, China}
\affiliation{Institute of Theoretical Physics, Chinese Academy of Sciences, Beijing 100190, China}

\author{Jin-Wei Dong}
\affiliation{Anhui Province Key Laboratory of Condensed Matter Physics at Extreme Conditions, High Magnetic Field Laboratory, Chinese Academy of Sciences, Hefei 230031, China}

\author{Ziqiang Wang}
\thanks{Corresponding Author: wangzi@bc.edu}
\affiliation{Department of Physics, Boston College, Chestnut Hill, MA 02467, USA}

\author{Sen Zhou}
\thanks{Corresponding Author: zhousen@itp.ac.cn}
\affiliation{Institute of Theoretical Physics, Chinese Academy of Sciences, Beijing 100190, China}
\affiliation{School of Physical Sciences, University of Chinese Academy of Sciences, Beijing 100049, China}

\begin{abstract}
We propose a symmetry-guided route to engineer tunable $p$-wave magnetism in a bilayer system composed of two AB-stacked antiferromagnetic square lattices with their collinear moments nonparallel to each other.
We show that an in-plane relative shift between the two layers selectively breaks the symmetries protecting spin degeneracy while preserving time-reversal-related constraints, thereby generating odd-parity spin splitting in a fully compensated magnetic state without spin-orbit coupling.
The direction and the magnitude of the resulting $p$-wave spin-splitting, together with the associated spin responses, can be continuously tuned by the in-plane displacement, providing a potential knob for control.
We further map the bilayer system onto an effective bond-modulated square lattice and investigate the corresponding Hubbard model within a mean-field framework.
The calculated phase diagram reveals extended regions where $p$-wave magnetism with coplanar or chiral spin textures emerges spontaneously.
Our results establish a minimal equilibrium platform for realizing and manipulating $p$-wave magnetism in two-dimensional systems.	
\end{abstract}
\maketitle

\textit{Introduction. -}
Unconventional magnetism has emerged as a transformative framework for next-generation spintronics \cite{vsmejkal2022emerging, bai2024altermagnetism, liu2025different,chen2026rise}.
A prominent class is spin-split antiferromagnetism (AFM), which combines the advantages of both ferromagnetism (FM) and AFM by hosting fully compensated magnetic moments yet exhibiting nonrelativistic spin splitting in the absence of spin-orbit coupling (SOC) \cite{jungwirth2026symmetry, jungwirth2025altermagnetism,liu2022spin, chen2024enumeration, xiao2024spin,wu2004dynamic}. 
The collinear manifestation of this class, formally termed as altermagnetism, has driven much of the recent progress in the field \cite{vsmejkal2022beyond, vsmejkal2022anomalous, vsmejkal2022emerging, bai2024altermagnetism, liu2025different, liu2022spin, chen2024enumeration, xiao2024spin, mazin2022altermagnetism, gonzalez2021efficient, vsmejkal2022giant, shao2021spin, roig2024minimal,  hoyer2025spontaneous, he2023nonrelativistic, zhou2024crystal, dai2024thermal, feng2022anomalous, bhowal2024ferroically, antonenko2025mirror, duan2025antiferroelectric, gu2025ferroelectric, takahashi2025elasto, jungwirth2026symmetry, jungwirth2025altermagnetism, song2025altermagnets, he2025altermagnetism, li2024d, chen2026rise, wu2004dynamic, yuan2026unconventional, yuan2026spin}.
Guided by symmetry analyses, numerous candidate altermagnetic materials have been theoretically predicted \cite{vsmejkal2020crystal, gonzalez2021efficient, mazin2021prediction, guo2023spin, krempasky2024altermagnetic, zhou2024crystal, liu2024chiral, lee2024broken, zhu2025design, parfenov2025pushing}, and experimental evidence for altermagnetic spin splitting has recently been reported in systems such as RuO$_2$, MnTe, and CrSb \cite{osumi2024observation, krempasky2024altermagnetic, feng2022anomalous, zeng2024observation, LiuY-NC25, liao2025direct, ding2024large, QianT-NP25}.
The collinearity of magnetic moments in altermagnetism imposes strictly even-parity spin-split band structures, such as $d$-wave or $g$-wave patterns \cite{wang2025spin, maier2023weak, brekke2023two, fernandes2024topological, das2024realizing, durrnagel2025altermagnetic, ferrari2024altermagnetism, leeb2024spontaneous, giuli2025altermagnetism, dong2025spontaneous, lu2026van}, in which the spin polarization is even under momentum inversion. 

Beyond even-parity altermagnetism of collinear magnets, recent theoretical and experimental studies have uncovered the possibility of odd-parity spin-split AFM in systems with noncollinear or noncoplanar magnetic order \cite{wu2004dynamic, hayami2022mechanism, Agterberg-PRL25, yuan2026unconventional, song2025electrical, Yamada-Nat25, brekke2024minimal, hellenes2023p, Lange-ASLG26}.
In such unconventional magnetic states, the spin polarization changes sign under momentum inversion, giving rise to $p$-wave and higher-order odd-parity spin textures that are distinct from altermagnetism.
In the symmetry classification, the $p$-wave class admits a lowest-order spin-polarization component linear in momentum, analogous to the orbital structure of $p$-wave superconductivity \cite{Hirsch-PRB90, WuCJ-PRB07}.
The odd-parity nature of $p$-wave magnetism offers promising opportunities for electrically and optically controllable, nonreciprocal, and potentially topological spintronic functionalities \cite{vsmejkal2022emerging, bai2024altermagnetism}.
Meanwhile, a growing range of emergent phenomena have been reported in such systems, including highly efficient nonrelativistic Edelstein effects \cite{chakraborty2025highly}, unconventional nonlinear optical responses \cite{sivianes2025optical}, purely electrical detection of the N\'{e}el vector via nonlinear conductivities \cite{ezawa2025purely}, and transverse spin-current generation at heterojunctions \cite{salehi2024transverse}. 
Furthermore, the interplay between $p$-wave magnetism and superconductivity has recently attracted increasing interest \cite{sukhachov2025coexistence, Law-PRB25, Fukaya-JPCM25}, opening new directions for superconducting spintronics.

Despite these advances, realizing robust and controllable $p$-wave magnetism remains challenging. 
On the theoretical side, symmetry-based classifications and minimal models have established $p$-wave magnetism as the odd-parity counterpart of altermagnetism and clarified its distinctive transport and topological responses \cite{hellenes2023p, brekke2024minimal, Agterberg-PRL25, Lange-ASLG26}. 
Experimentally, signatures of $p$-wave spin splitting have been reported in systems with complex noncollinear spin textures, such as helimagnetic metals \cite{song2025electrical, Yamada-Nat25}. 
However, these realizations generally rely on intricate magnetic orders and material-specific conditions, limiting their tunability and generality. 
More recently, alternative equilibrium mechanisms have been proposed based on itinerant and quantum-spin models, orbital-current order, orbital magnetism, and engineered heterostructures \cite{Lee-PRL25, GiBaik-PRL26, LinYP-PRL26, Leeb-PRL26, Li-arXiv26, Jin-arXiv26, Kim-PRB26}, some of which offer electrical control through carrier filling or gating.
Nonequilibrium schemes based on polarized laser pumping provide another means of dynamically inducing odd-parity magnetic responses through light-matter coupling and inverse Faraday effects \cite{YanZB-PRB26, WangR-PRL26, RuanJW-PRL26, Kovalev-PRL26}, but require continuous external driving. 
These considerations highlight the need for simple equilibrium platforms in which $p$-wave magnetism can be intrinsically realized and continuously controlled through a direct geometric degree of freedom.

In this work, we propose a symmetry-transparent route to engineer $p$-wave magnetism based on two AB-stacked AFM square lattices with their collinear magnetic moments nonparallel to each other.
Upon introducing an in-plane relative shift, we demonstrate that an odd-parity $p$-wave magnetic state naturally emerges from symmetry considerations. 
Remarkably, the direction of the $p$-wave order, the strength of the spin splitting, and the resulting spin responses can all be continuously tuned by the in-plane shift, providing a potential knob for control.
We further show that this bilayer configuration maps onto an effective bond-modulated square lattice, in which the nearest-neighbor (NN) hoppings are anisotropically modulated while the next-NN (NNN) hoppings remain uniform. 
We then investigate the corresponding Hubbard model within a mean-field framework and determine the magnetic phase diagram over a broad parameter regime. 
We find that $p$-wave magnetism with either coplanar or chiral spin configurations emerges spontaneously and is energetically favored in an extended region of the parameter space, though further beyond-mean-field studies are required to establish its stability against quantum fluctuations.
Our results therefore establish geometric displacement as a generic symmetry-control parameter capable of selectively breaking spin-space symmetries and converting compensated AFM into $p$-wave magnetic states.
This mechanism provides a simple equilibrium route toward tunable $p$-wave magnetism in two-dimensional systems, without requiring complex spin textures, strong SOC, or external periodic driving.

\begin{figure}
\begin{center}
\fig{3.4in}{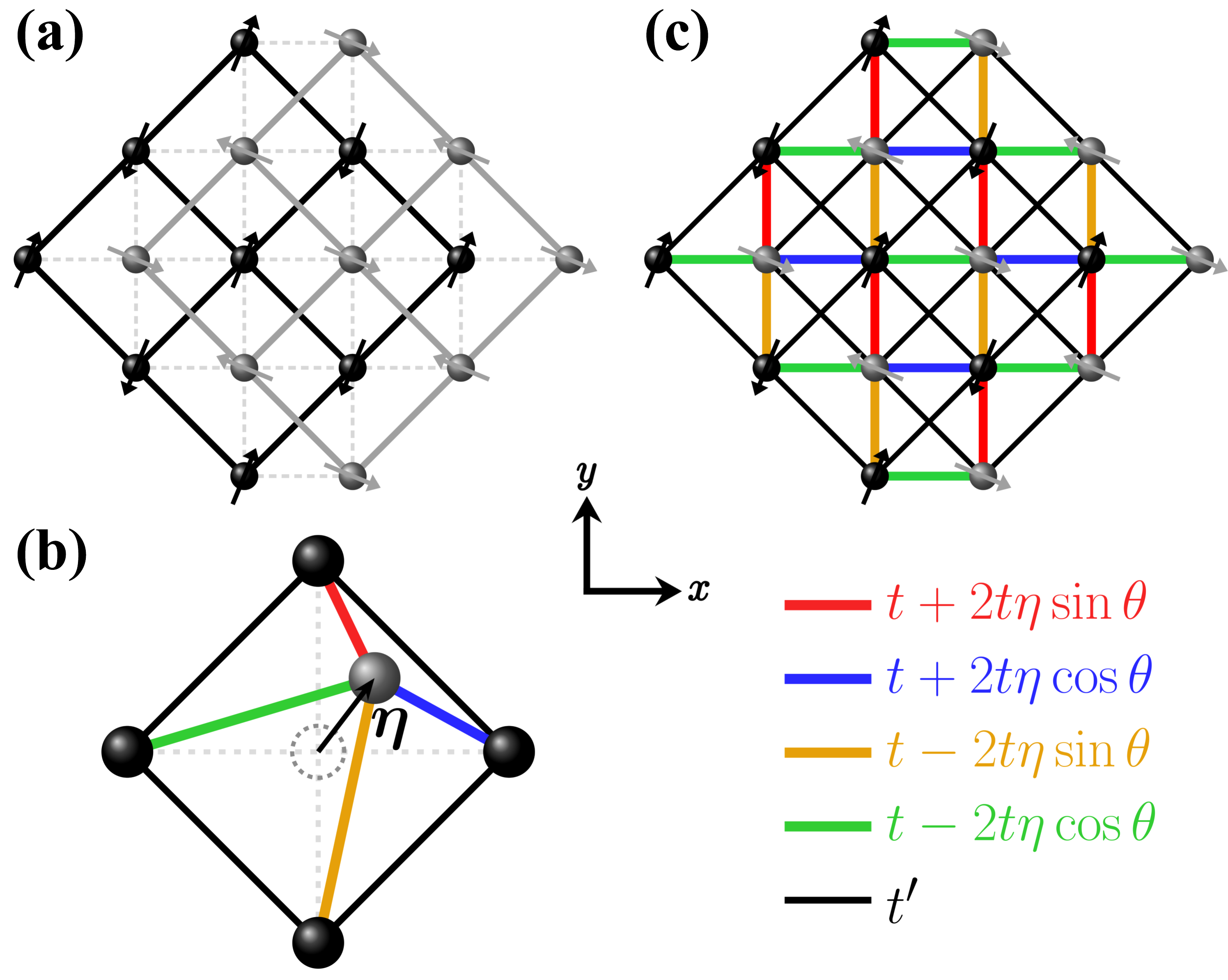}
\caption{(a) Top view of the two AB-stacked, AFM-ordered square lattices with mutually perpendicular collinear moments in the $xy$-plane. (b) In-plane shift ${\bm \eta}$ = $\eta(\cos\theta, \sin\theta)$ and the induced anisotropic modulation in the interlayer NN hoppings. (c) The bond-modulated square lattice with anisotropically modulated NN hoppings and uniform NNN hoppings. Inset defines the $xy$ coordinate.} \label{fig1}
\end{center}
\vskip-0.6cm
\end{figure}

\textit{Spin space group symmetry analysis.} - 
We begin by considering two AB-stacked, AFM-ordered square lattices with their collinear moments nonparallel to each other and, without loss of generality, lying in the $xy$-plane.
The detailed spin space group symmetry analyses in the absence of SOC are provided in the Supplementary Material (SM) \cite{supp}, where $\{\mathcal{O}_S||\mathcal{O}_R {\bm \tau}\}$ denotes a unitary spin-space operation with $\mathcal{O}_S$ and $\mathcal{O}_R {\bm \tau}$ acting on, respectively, the spin space and the real space, with the spatial translation $\bm \tau$ retained whenever applicable.
Here, we focus on the perpendicular configuration displayed in Fig. \ref{fig1}(a), which represents a symmetry-transparent limit of the nonparallel condition.
For spins along the $z$-axis, the even-parity relation $E(\bk,s_z)$ = $E(-\bk,s_z)$ is independently protected by the spatial two-fold rotation $C_{2z}$ and the combined symmetry $\{C_{2z}||C_{2z}{\bm\tau}_3\}$, where ${\bm \tau}_3$ represents the elementary translation connecting the opposite magnetic sublattices and the rotation axis of spatial $C_{2z}$ passes a lattice site, \textit{i.e.}, any black or gray dot in Fig. \ref{fig1}(a).
At the same time, the odd-parity relation $E(\bk,s_z)$ = $E(-\bk,s_{-z})$ is independently enforced by the effective time-reversal $\mT {\bm \tau}_3$ and the combined symmetry $\{C_{2z}||\mT\}$.
Consequently, the spin degeneracy along $z$-direction is strictly protected at every momentum, $E(\bk,s_z)$ = $E(\bk,s_{-z})$.
Similarly, for spins along any in-plane direction $z_\perp$ in the $xy$-plane, the even-parity relation $E(\bk,s_{z_\perp})$ = $E(-\bk,s_{z_\perp})$ is protected independently by spatial rotation $C_{2z}$ and $\{C_{2z}||\mT\}$, while the odd-parity relation $E(\bk,s_{z_\perp})$ = $E(-\bk,s_{-z_\perp})$ is independently imposed by $\mT {\bm \tau}_3$ and $\{C_{2z}||C_{2z} {\bm \tau}_3\}$, resulting in the spin degeneracy $E(\bk,s_{z_\perp})$ = $E(\bk,s_{-{z_\perp}})$ throughout the momentum space.
Therefore, despite the noncollinear arrangement of the magnetic moments, the system shown in Fig. \ref{fig1}(a) remains spin degenerate in all directions.

\begin{figure*}
\begin{center}
\fig{7.in}{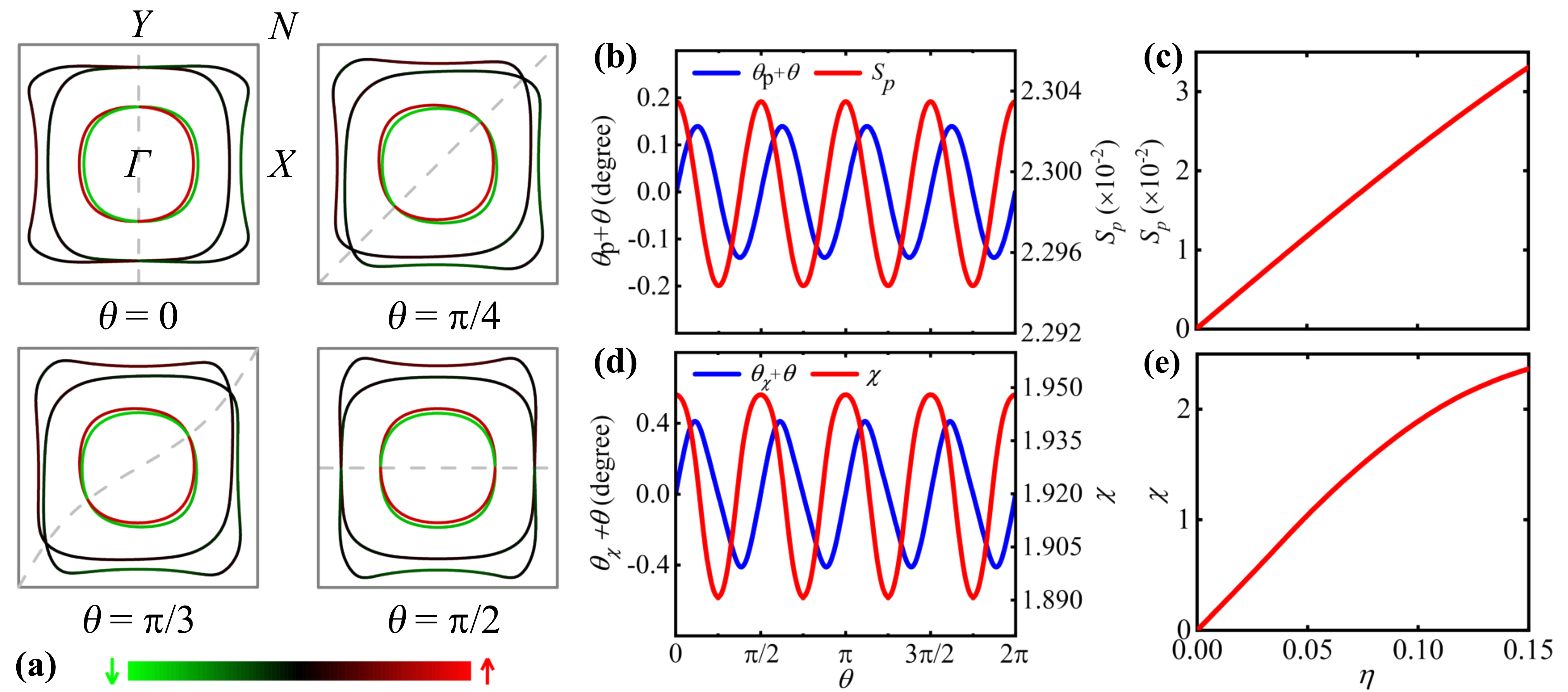}
\vskip-0.1cm
\caption{(a) Equal-energy contours at $E=-t$ for representative shift angles, $\theta$ = 0, $\pi/4$, $\pi/3$, and $\pi/2$ at fixed shift magnitude $\eta$ = 0.1, where line color denotes the spin polarization with red and green for $s^z_{n\bk} = \pm 1$.
The dashed gray curves display the spin-degeneracy nodal lines and the gray square represents the $2\times 2$ reduced BZ, with high-symmetry points $\Gamma =(0, 0)$, $X=(\pi/2, 0)$, $Y=(0, \pi/2)$, and $N=(\pi/2, \pi/2)$. (b) The direction of the $p$-wave order $\theta_p$ and the strength of $p$-wave spin splitting as a function of shift angle $\theta$ at fixed shift magnitude $\eta$ = 0.1. (c) The strength $S_p$ as a function of $\eta$ at fixed $\theta=\pi/4$.
The system is half-filled in (a-c). (d) The direction $\theta_\chi$ and the magnitude $\chi$ of the response vector $\bm \chi$ as a function of shift angle $\theta$ at fixed shift amplitude $\eta$ =0.1. (e) The magnitude $\chi$ as a function of $\eta$ at fixed $\theta=\pi/4$.
In (d) and (e), the filling is 0.9 electrons per site and the magnitude $\chi$ is measured in units of $e\hbar/ta_0$.} \label{fig2}
\end{center}
\vskip-0.6cm
\end{figure*}

Interestingly, an arbitrary in-plane shift ${\bm \eta}$ shown in Fig. \ref{fig1}(b) breaks both the spatial rotation $C_{2z}$ and the combined symmetry $\{C_{2z}||C_{2z} {\bm\tau}_3\}$, while preserving the effective time-reversal $\mT {\bm\tau}_3$ and the combined symmetry $\{C_{2z}||\mT\}$.
This selective symmetry breaking naturally allows the emergence of odd-parity spin polarization along the $z$-direction, normal to the plane of magnetic moments, while maintaining the spin degeneracy for all in-plane direction $z_\perp$. 
Specifically, it belongs to the anti-spin Laue group $^2{\bar 1}$ with halving subgroup $H$ = 1, which is classified as a $p$-wave anti-altermagnet \cite{Lange-ASLG26}.
Here and throughout this work, we use $p$-wave magnetism in this symmetry sense; the full spin polarization may additionally contain symmetry-allowed higher-order odd-parity lattice harmonics.
Moreover, the fully compensated zero net magnetization of the system is guaranteed by the $\{C_{2z}||{\bm \tau}_3\}$ symmetry in the presence of a finite shift $\bm \eta$.
Therefore, we show that a $p$-wave magnetism can be realized by introducing an in-plane shift in two AB-stacked, AFM-ordered square lattices with mutually perpendicular collinear moments.

Explicitly, an in-plane shift along the $x$-axis breaks the mirror symmetry $M_x$ while preserving $M_y$, thereby inducing a $p_x$-wave magnetism.
Likewise, a shift along the $y$-axis gives rise to a $p_y$-wave magnetism.
Consequently, a general in-plane shift ${\bm \eta}$ = $\eta(\cos\theta, \sin\theta)$ shown in Fig. \ref{fig1}(b) generates a combined $(p_x, p_y)$ = $p(\cos\theta_p, \sin\theta_p)$-wave magnetism, where the angle $\theta_p$ characterizes the direction of the $p$-wave order.
Landau free energy analysis provided in SM \cite{supp} shows that, to quadratic order, the $p_x$- and $p_y$-wave components induced by shifts along the $x$- and $y$-directions differ by a relative $\pi$ phase.
As a result, the direction of the emergent $p$-wave order follows that of the in-plane shift, with $\theta_p$ = $-\theta$ or $\theta_p$ = $\pi -\theta$ being satisfied exactly along high-symmetry directions and exhibiting slight deviations elsewhere.
Furthermore, it is natural to expect that the strength of $p$-wave spin splitting is controlled by the magnitude of shift.
Therefore, both the direction of the $p$-wave order and the strength of the spin splitting can be continuously tuned by the in-plane shift, potentially providing a control knob for spintronic applications.
Although the perpendicular spin configuration in Fig. \ref{fig1}(a) considered here provides the simplest symmetry realization, the emergence of $p$-wave spin splitting generally requires only that the collinear AFM moments in the two layers are nonparallel \cite{supp}.

\textit{Phenomenological modeling.} - 
To validate the symmetry analysis and demonstrate the realization and manipulation of $p$-wave magnetism through an in-plane shift, we consider a phenomenological model for bilayer systems presented in Figs. \ref{fig1}(a) and \ref{fig1}(b).
The in-plane shift $\bm \eta$ effectively modulates anisotropically the inter-layer NN hoppings while leaving the intra-layer NN hoppings unchanged, as shown schematically in Fig. \ref{fig1}(b).
Consequently, the bilayer system can be mapped onto the bond-modulated square lattice displayed in Fig. \ref{fig1}(c), where the NN hoppings modulate anisotropically while the NNN hoppings remain uniform.
The corresponding phenomenological Hamiltonian is therefore given by
\begin{equation}
H_p = -\sum_{ij,\alpha} t_{ij}c^\dagger_{i\alpha} c_{j\alpha} - h\sum_i {\bm e}_i \cdot \hat{\bm m}_i, \label{Hp}
\end{equation}
where $c^\dagger_{i\alpha}$ creates a spin-$\alpha$ electron at site $i$, and the nonzero hoppings $t_{ij}$ are illustrated in Fig. \ref{fig1}(c). 
We set the averaged NN hopping $t=1$ as the energy unit and fix the NNN hopping $t'=1.2t$.
The phenomenological local magnetic field has magnitude $h=1.5t$ and direction ${\bm e}_i$, which pins the magnetic moment configuration displayed in Fig. \ref{fig1}(c).
The local moment operator $\hat{\bm m}_i$ = $\sum_{\alpha \beta} c^\dagger_{i\alpha} {\bm \sigma}_{\alpha\beta} c_{i\beta}$, where ${\bm \sigma}$ = $(\sigma_x, \sigma_y, \sigma_z)$ is the vector of spin Pauli matrices.
Eq. (\ref{Hp}) is then transformed into momentum space and diagonalized to obtain the spin-resolved electronic structure, which exhibits band- and momentum-dependent spin splitting along $z$-axis \cite{supp}.

We stay at half-filling with one electron per site. 
First, we examine the state evolution as a function of shift angle $\theta$ at a fixed magnitude $\eta=0.1$.
The spin-resolved equal-energy contours at $E=-t$ are displayed in Fig. \ref{fig2}(a) for representative angles, $\theta$ = 0, $\pi/4$, $\pi/3$, and $\pi/2$.
While the outer pair of contours remain largely spin degenerate, the inner pair of contours exhibit clear and pronounced $p$-wave spin splitting, with the orientation of the $p$-wave order continuously evolving as $\theta$ varies.
The nodal line shown in Fig. \ref{fig2}(a), on which the spin splitting vanishes in all bands, further illustrates the $p$-wave character of the calculated spectrum.
To characterize quantitatively the unconventional magnetic state, we introduce a vector order parameter for the $p$-wave spin splitting
\begin{equation}
\textbf{S}_p = \frac{1}{N} \sum_\bk s^z(\bk) {\bm \phi}_\bk, \hspace{0.3cm} s^z(\bk) = \sum_n s^z_{n\bk} f(E_{n\bk}), \label{Sp}
\end{equation}
where $N$ is the number of sampled \textbf{k} points, $s^z_{n\bk}$ = $\langle u_{n\bk}| \sigma_z |u_{n\bk}\rangle$ denotes the dimensionless $z$-axis spin polarization of band $n$ at momentum \bk, with $|u_{n\bk} \rangle$ the periodic part of the Bloch state wave function.
The corresponding band dispersion $E_{n\bk}$ is measured relative to the Fermi level, and $f$ denotes the Fermi-Dirac distribution function.
The form factor ${\bm \phi}_\bk = (\sin k_x, \sin k_y)$ extracts the lowest-harmonic $p$-wave component of spin polarization of all occupied-states, while higher odd lattice harmonics are not excluded.

The vector order parameter can be expressed as $\textbf{S}_p$ = $S_p (\cos\theta_p, \sin\theta_p)$, where the amplitude $S_p$ characterizes the strength of $p$-wave spin splitting and the angle $\theta_p$ specifies the direction of the $p$-wave order.
They are presented in Fig. \ref{fig2}(b) as a function of shift angle $\theta$ at fixed amplitude $\eta=0.1$.
Clearly, the direction of the $p$-wave order follows roughly $\theta_p$ = $-\theta$, consistent with the Landau free energy analysis.
Reversing the local magnetic field on one of the two layers changes the relation to $\theta_p=\pi-\theta$.
Meanwhile, the strength of $p$-wave spin splitting $S_p$ is nearly independent of $\theta$, with a weak $\pi/2$-periodic oscillation reflecting the $C_{4v}$ symmetry of the unshifted bilayer \cite{supp}.
We then vary the shift magnitude $\eta$ at a fixed angle $\theta=\pi/4$. 
The strength of $p$-wave spin splitting $S_p$ vanishes at $\eta=0$, implying a spin-degenerate electronic structure in the absence of in-plane shift.
Upon increasing $\eta$, a $p$-wave magnetic state emerges with finite spin-splitting strength $S_p$ plotted in Fig. \ref{fig2}(c).
Clearly, it grows almost linearly with $\eta$ up to $\eta = 0.15$.
We note that the dependence of $S_p$ on $\eta$ deviates from linearity at large $\eta$ and may even become nonmonotonic when the system is doped away from half-filling \cite{supp}.
Nevertheless, the direction of the $p$-wave order, as well as the strength of the spin splitting, can be tuned continuously through the in-plane shift $\bm \eta$, providing a practical control of the $p$-wave magnetic state.

In the absence of SOC, the odd-parity spin splitting of $p$-wave magnetism, together with the symmetries of the bilayer system, forbids any spin Hall response while allowing a finite Edelstein effect.
Namely, an external electric field can induce a nonzero spin-polarization density, $\delta S^a$ = $\sum_b \chi^{ab} E_b$, where the charge-to-spin conversion tensor is given by
\begin{equation}
\chi^{ab} = \frac{e\hbar}{4N\Gamma A} \sum_{n,\bk} \langle u_{n\bk}| \sigma_a |u_{n\bk} \rangle \langle u_{n \bk} | {\partial H(\bk)} / {\partial k_b} |u_{n\bk}\rangle \delta(E_{n\bk}), \label{chi}
\end{equation}
within the linear response theory.
Here, $A = 4a_0^2$ denotes the area of the $2\times 2$ enlarged unit cell of the bond-modulated square lattice, with $a_0$ the pristine lattice constant.
The delta function is approximated by $\delta(E) =\tfrac{1}{\pi} \tfrac{\Gamma} {E^2+\Gamma^2}$, with constant scattering rate $\Gamma = 0.01t$.
Eq. (\ref{chi}) contains the intra-band contributions associated with the Fermi surface shift under an external electric field, while the inter-band contributions vanish in the presence of $\mathcal{T}{\bm \tau}_3$ symmetry.
For the present $p$-wave magnetic state with spin splitting along the $z$-axis, the nonzero elements of the response tensor are $\chi^{zx}$ and $\chi^{zy}$, from which we define a response vector ${\bm \chi} =(\chi^{zx}, \chi^{zy}) \equiv \chi (\cos \theta_\chi, \sin \theta_\chi)$.

Here, we move to 0.9 electron filling where the system is metallic with well-defined Fermi surface, which is required for a nonzero Edelstein effect.
First, we vary the shift angle $\theta$ at a fixed amplitude $\eta$ = 0.1. 
The angle $\theta_\chi$ and the magnitude $\chi$ of the response vector are displayed in Fig. \ref{fig2}(d) as a function of $\theta$.
Again, the direction of $\bm \chi$ follows that of the in-plane shift with roughly $\theta_\chi = -\theta$, and the magnitude $\chi$ remains nearly constant except for a weak $\pi/2$-periodic oscillation, similar to the behavior of $S_p$.
Using the lattice constant $a_0$ = 4 \AA\ and the NN hopping $t$ = 0.3 eV, the estimated magnitude of Edelstein response $\chi \simeq 1.58 \hbar/\text{V\AA}$, indicating a potentially substantial charge-to-spin conversion response \cite{chakraborty2025highly, edelstein1990spin, johansson2016theoretical, manchon2015new}.
We then vary the shift amplitude $\eta$ at a fixed angle $\theta=\pi/4$. 
The response magnitude $\chi$ increases monotonically as a function of $\eta$, as displayed in Fig. \ref{fig2}(e).
We note that qualitatively similar behaviors are reproduced in phenomenological calculations with generic nonparallel magnetic configurations \cite{supp}.
These results thus validate the symmetry analyses and clearly demonstrate that the in-plane shift provides an efficient geometric knob to realize and manipulate $p$-wave magnetism and the associated spin response in AB-stacked, AFM-ordered square lattices without relying on SOC.

\begin{figure}
	\begin{center}
		\fig{3.4in}{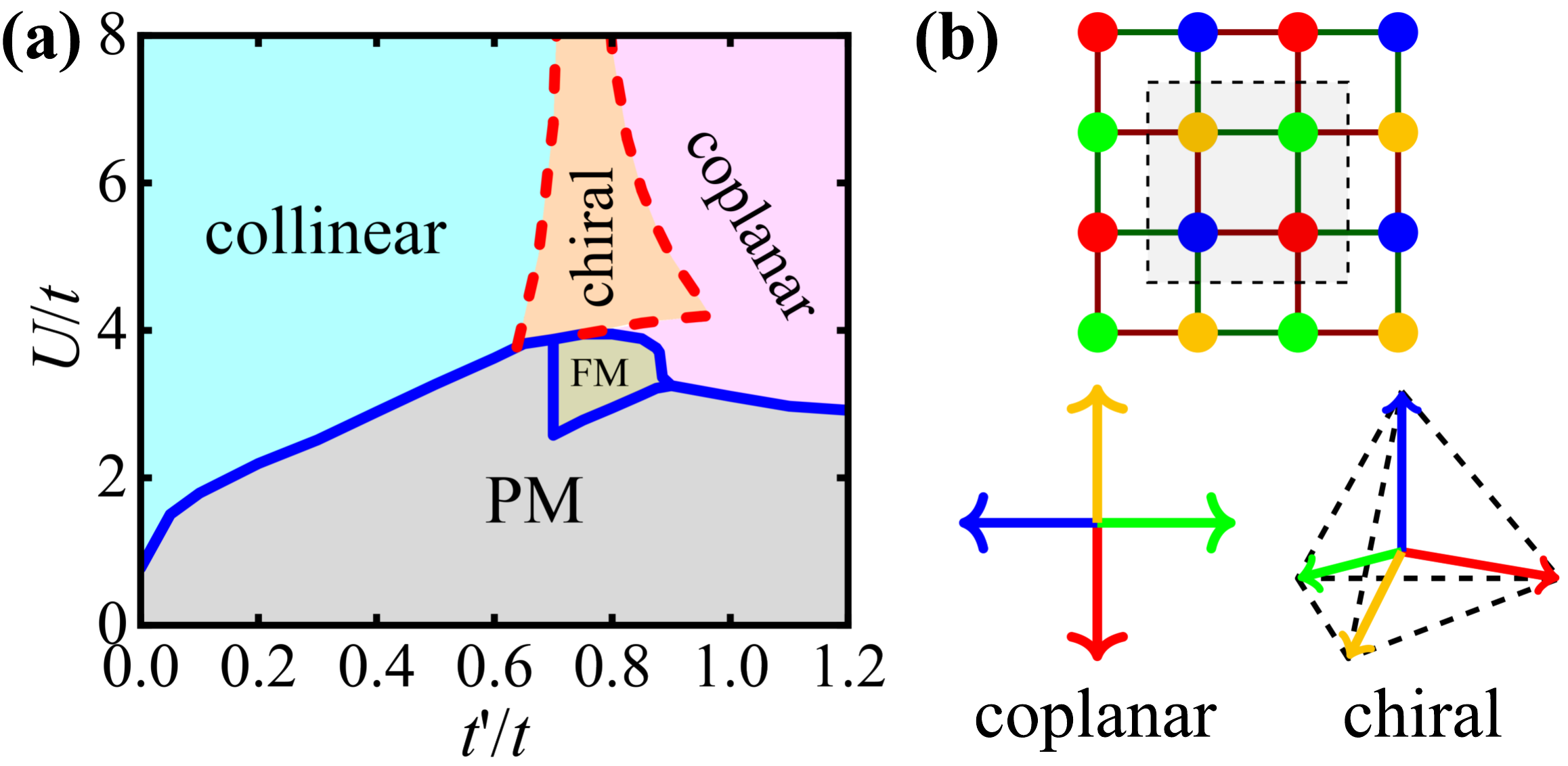}
		\vskip-0.1cm
		\caption{(a) The mean-field magnetic phase diagram of the half-filled Hubbard model on the bond-modulated square lattice associated with in-plane shift $(\eta, \theta)$ = $(\sqrt{2}/10, \pi/4)$. Solid and dashed lines denote, respectively, first-order and continuous phase transitions. (b) The $2\times2$ enlarged unit cell of coplanar and chiral AFM phases, in which the magnetic moments on the four inequivalent sites form, respectively, a cross and a tetrahedron in spin space.} \label{fig3}
	\end{center}
	\vskip-0.6cm
\end{figure}

\textit{Microscopic model realization.} -  
The preceding symmetry analysis assumes nonparallel AFM moments, which the phenomenological model fixes with local fields.
We now ask whether on-site Coulomb repulsion can spontaneously generate $p$-wave magnetic order in the shifted bilayer.
To this end, we study the Hubbard model on the corresponding bond-modulated square lattice shown in Fig. \ref{fig1}(c),
\begin{equation}
H = -\sum_{ij,\sigma} t_{ij}c^\dagger_{i\sigma} c_{j\sigma} + U\sum_i \hat{n}_{i\uparrow} \hat{n}_{i\downarrow}, \label{Hubbard}
\end{equation}
where Hubbard $U$ is treated within the SU(2) spin-rotation-invariant Hartree-Fock approximation.
The resulting mean-field Hamiltonian reads
\begin{equation}
H_\text{MF} = -\sum_{ij,\sigma} t_{ij}c^\dagger_{i\sigma} c_{j\sigma} + \frac{U}{4} \sum_i \left[ 2n_i \hat{n}_i -n^2_i -2{\bm m}_i \cdot \hat{\bm m}_i +{\bm m}^2_i \right], \nonumber
\end{equation}
where the order parameters $n_i$ = $\langle \hat{n}_i \rangle$ and ${\bm m}_i$ = $\langle \hat{\bm m}_i \rangle$ are determined self-consistently by minimizing the state energy.
We consider the quantum states restricted to commensurate configurations compatible with $2\times 2$ enlarged unit cell, and use different initial conditions and compare the energies of the converged solutions to identify the lowest-energy state obtained within the restricted space.
Incommensurate spiral and spin-density-wave states are not included.
Such orders may compete with commensurate magnetism \cite{LeeCH-PRB26}, and their possible occurrence in the present model remains to be investigated.
All the low-energy states obtained here are phases with uniform charge density, as the onsite repulsion $U$ suppresses charge fluctuations.

We consider the bond-modulated square lattice corresponding to the bilayer system with in-plane shift $\eta$ = $\sqrt{2}/10$ and $\theta$ = $\pi/4$.
Fig. \ref{fig3}(a) displays the mean-field magnetic phase diagram at half-filling in the plane spanned by $U$ and $t'$, which is very similar to the one obtained in the absence of bond modulation \cite{huang2020antiferromagnetic}.
The paramagnetic (PM) phase at small $U$ is replaced by magnetically ordered states at all $t'/t$ ratios when the interaction $U$ is sufficiently strong.
Except for a small FM region near moderate $U\simeq 3t$ and $t'\simeq 0.8t$, the magnetic states are all AFM with zero net magnetization.
The AFM state in a narrow regime near $t'\simeq 0.8t$ is a triple-$\bQ$ chiral AFM, sandwiched by the conventional N\'{e}el AFM phase with collinear moments at small $t'$ and the double-$\bQ$ coplanar AFM at large $t'$.
As illustrated in Fig. \ref{fig3}(b), the four sublattice moments in the $2\times 2$ enlarged unit cell lie in a common plane and form a cross in the coplanar AFM phase, giving zero scalar spin chirality.
In the chiral AFM phase, they form a noncoplanar tetrahedron in spin space with finite scalar spin chirality.
We note that the chiral AFM order can be viewed as a superposition of a collinear AFM order and a coplanar AFM order orthogonal to it. 
Furthermore, the Hubbard model is invariant under spin rotation.
We thus orient, for convenience, the collinear AFM order to the $z$-direction and the coplanar AFM order within the $xy$-plane.

The spin-resolved electronic structures of these magnetic states are presented in the SM \cite{supp}.
Consistent with symmetry analysis and phenomenological calculations, both the coplanar AFM and the chiral AFM exhibit unconventional band- and momentum-dependent $p$-wave spin splitting exclusively along $z$-direction. 
Interestingly, the finite spin chirality breaks both parity and time-reversal symmetry, leading to topologically nontrivial electronic bands \cite{supp}.
Consequently, doping the chiral AFM phase away from half-filling may simultaneously give rise to anomalous Hall and Edelstein effects.
Additional results on bond-modulated square lattices corresponding to different in-plane shifts are provided in SM \cite{supp}.
It is further illustrated that the in-plane shift provides continuous control of the direction and the magnitude of the $p$-wave spin splitting.
Interestingly, when the in-plane shift is not along the diagonal directions, the magnetic moments in the coplanar $p$-wave magnetism are not exactly perpendicular to each other, justifying that the emergence of $p$-wave magnetic states requires only nonparallel moments.
Furthermore, the coplanar and chiral $p$-wave magnetic states persist over finite doping ranges.
These results demonstrate that unconventional $p$-wave magnetism can emerge spontaneously in the Hubbard model on bond-modulated square lattices and, equivalently, the shifted bilayer square lattices, and is energetically favored among the restricted commensurate mean-field solutions in an extended parameter region.
Further beyond-mean-field studies are required to establish the stability of these states against quantum fluctuations.

\textit{Summary and discussions.} -
In this work, we have theoretically proposed a plausible route to engineer unconventional $p$-wave magnetism by introducing an in-plane relative shift in a bilayer system composed of two AB-stacked, AFM-ordered square lattices with nonparallel collinear moments. 
Symmetry analysis shows the fully compensated zero net magnetization is protected by spin-space symmetries that map spins onto opposite partners.
The $p$-wave magnetism emerges from selectively breaking the symmetries protecting spin degeneracy, while preserving the time-reversal-related constraints responsible for odd-parity spin splitting.
Supported by phenomenological modeling, we demonstrate that the direction of the $p$-wave order, the magnitude of the spin splitting, and the associated spin responses can all be continuously tuned through the in-plane shift, providing an efficient control parameter for spintronic applications.

A possible experimental realization of our proposal may be pursued in oxide heterostructures composed of layered AFM such as Sr$_2$IrO$_4$ or La$_2$CuO$_4$ \cite{KimBJ-science09, Keimer-PRB92}, both of which host quasi-two-dimensional square lattices.
Modern epitaxial growth techniques allow the fabrication of bilayer structures with controlled stacking geometries and interface properties \cite{Hwang-NM12, Chakhalian-RMP14}. 
A nonparallel alignment of the two N\'{e}el vectors may be stabilized by competition between interlayer and layer-dependent magnetic anisotropies with different easy axes, induced, for example, by strain, octahedral rotations, or inequivalent interfaces \cite{HeJ-PRL10, LuW-SR15, YiD-PRL17}. 
In addition, an in-plane relative shift between the two layers can be introduced through substrate-induced lattice mismatch, controlled shear strain, or deliberately designed stacking offsets during growth \cite{Hwang-NM12, Zubko-ARCMP11}. 
Beyond oxide heterostructures, van der Waals AFM bilayers may provide an alternative and highly tunable platform, where structural degrees of freedom such as relative layer displacement, stacking geometry, and interlayer registry can be continuously controlled and have been shown to strongly influence magnetic ordering and symmetry properties \cite{CaoY-Nat18, Soriano-NL20, LiT-NM19}.
Recent experiments on bilayer boron nitride have demonstrated reversible switching of out-of-plane electric polarization through electrically driven interlayer sliding, providing an experimental proof of principle for controlled relative shift in van der Waals bilayers \cite{Yasuda-Science24}.
Interface engineering has been demonstrated to modify magnetic exchange interactions and lattice symmetries \cite{Chakhalian-RMP14, YiD-PRL17}, providing a possible route to combine the key ingredients of our proposal — square-lattice AFM layers, nonparallel N\'{e}el vectors, and a controllable in-plane shift — in a layered magnetic platform.
We note that a concrete material realization of our proposal will require a material-specific analysis of the atomic registry, interface termination, and magnetic configuration \cite{Sasioglu-PRB26, Lange-PRX26}.

The shifted bilayer system considered here maps onto a bond-modulated square lattice.
Mean-field calculations in $2\times 2$ variational spaces indicate that $p$-wave magnetism can emerge spontaneously from sufficiently strong on-site Coulomb interactions at large $t'/t$ ratio, with either coplanar or chiral spin configurations.
Such large $t'/t$ ratios may be realized naturally by adjusting the interlayer distance between the two square-lattice layers.
Nevertheless, we note that more accurate many-body studies of the square-lattice Hubbard model without bond modulation, including DMRG and QMC calculations, generally favor stripe-like collinear magnetic orders rather than multi-\bQ\ states, even at large $t'$ \cite{Huang-npj18, QinMP-ARCMP22}.
This suggests that, although multiple ordering vectors may be nearly degenerate, quantum fluctuations suppress the phase coherence required to stabilize coplanar or chiral AFM states.
Interestingly, the same bond modulation that enables $p$-wave splitting in the coplanar state may also reshape higher-order magnetic interactions that select among stripe and multi-\bQ\ states, in analogy with ring-exchange and effective biquadratic mechanisms \cite{Hayami-PRB17, Larsen-PRB19, Hayami-JMMM20}.
It is therefore desirable to investigate the bond-modulated square lattice Hubbard model using more accurate many-body approaches in future studies.

\textit{Acknowledgments.} -  
This work is supported by the National Key Research and Development Program of China (Grant No. 2022YFA1403800), the National Natural Science Foundation of China (Grant Nos. 12374153 and 12447101), and the Postdoctoral Fellowship Program of CPSF (Grant No. GZC20241749).
Z.W. is supported by the U.S. Department of Energy, Basic Energy Sciences
(Grant No. DE-FG02-99ER45747).
Numerical calculations in this work were performed on the HPC Cluster of ITP-CAS.

\bibliography{bib}

\clearpage

\end{document}